\documentclass[journal,10pt]{IEEEtran}

\usepackage{color}
\usepackage{amsmath,amssymb,amsfonts}
\usepackage{graphicx}
\usepackage{verbatim}

\usepackage[font=scriptsize,caption=false,labelsep=space]{subfig}

\usepackage{url}
\usepackage[para,online,flushleft]{threeparttable}

\usepackage{booktabs}
\usepackage{multicol}

\usepackage{enumitem}

\usepackage[switch]{lineno}
\usepackage[named]{algo}
\usepackage[noend]{algpseudocode}
\usepackage{algorithm}

\usepackage{cite}
\makeatletter
\newcommand{\citecomment}[2][]{\citen{#2}#1\citevar}
\newcommand{\citeone}[1]{\citecomment{#1}}
\newcommand{\citetwo}[2][]{\citecomment[,~#1]{#2}}
\newcommand{\citevar}{\@ifnextchar\bgroup{;~\citeone}{\@ifnextchar[{;~\citetwo}{]}}}
\newcommand{\citefirst}{\@ifnextchar\bgroup{\citeone}{\@ifnextchar[{\citetwo}{]}}}
\newcommand{\cites}{[\citefirst}
\makeatother

\usepackage{balance}

\usepackage{amsmath}
\usepackage{amsthm}
\usepackage{dsfont}
\usepackage{thmtools}
\usepackage{amssymb}
\usepackage[dvipsnames]{xcolor}

\def\be{\boldsymbol{e}}

\def\bC{\boldsymbol{C}}

\def\balpha{\boldsymbol{\alpha}}

\def\bzeta{\boldsymbol{\zeta}}

\def\bmu{\boldsymbol{\mu}}
\def\bnu{\boldsymbol{\nu}}

\def\bPhi{\boldsymbol{\Phi}}

\def\mba{\mathbf{a}}
\def\mbb{\mathbf{b}}

\def\mbh{\mathbf{h}}

\def\mbp{\mathbf{p}}

\def\mbv{\mathbf{v}}
\def\mbw{\mathbf{w}}
\def\mbx{\mathbf{x}}
\def\mby{\mathbf{y}}

\def\mbA{\mathbf{A}}
\def\mbB{\mathbf{B}}

\def\mbE{\mathbf{E}}
\def\mbF{\mathbf{F}}
\def\mbG{\mathbf{G}}

\def\mbI{\mathbf{I}}

\def\mbP{\mathbf{P}}

\def\mbR{\mathbf{R}}
\def\mbS{\mathbf{S}}

\def\mbZ{\mathbf{Z}}

\def\bzero{\boldsymbol{0}}

\def\Tr#1{\mathrm{Tr}\left(#1\right)}

\def\diag#1{\mathrm{diag}\left(#1\right)}
\def\Diag#1{\mathrm{Diag}\left(#1\right)}

\def\Re#1{\operatorname{Re}\left(#1\right)}
\def\Im#1{\operatorname{Im}\left(#1\right)}
\theoremstyle{definition}

\def\LoS{{\textrm{LoS}}}
\def\NLoS{{\textrm{NLoS}}}

\def\T{\top}
\def\H{\mathrm{H}}

\makeatletter
\newcommand*{\rom}[1]{\expandafter\@slowromancap\romannumeral #1@}
\makeatother

\newcommand{\ze}[1]{\textcolor{black}{#1}}

\usepackage[english]{babel}

\newtheorem{theo}{Theorem}

\begin{document}
\title{Cram\'{e}r–Rao Lower Bound Optimization for Hidden Moving Target Sensing via Multi-IRS-Aided Radar}
%\title{Doppler-resilient Multi-IRS-Aided  radar  design based on  Cramér–Rao Lower Bound Optimization}
\author{Zahra Esmaeilbeig, Kumar Vijay Mishra, Arian Eamaz and Mojtaba Soltanalian
\thanks{Zahra Esmaeilbeig, Arian Eamaz and Mojtaba Soltanalian are with the ECE Departement, University of Illinois at Chicago, Chicago, IL 60607 USA. Email: \{zesmae2, aeamaz2, msol\}@uic.edu.}
\thanks{Kumar Vijay Mishra is with the United States DEVCOM Army Research Laboratory, Adelphi, MD 20783 USA. E-mail: kvm@ieee.org.}
\thanks{This work was sponsored in part by the National Science Foundation Grant  ECCS-1809225, and in part by the Army Research Office, accomplished under Grant Number W911NF-22-1-0263. The views and conclusions contained in this document are those of the authors and should not be interpreted as representing the official policies, either expressed or implied, of the Army Research Office or the U.S. Government. The U.S. Government is authorized to reproduce and distribute reprints for
Government purposes notwithstanding any copyright notation herein.}
%\thanks{The conference precursor of this work was presented at the 2022 IEEE Sensor Array and Multichannel Signal Processing Workshop (SAM).}	
}
\markboth{
}
{Shell \MakeLowercase{\textit{et al.}}: Bare Demo of IEEEtran.cls for IEEE Journals}
\maketitle

\begin{abstract}
Intelligent reflecting surface (IRS) is a rapidly emerging paradigm to enable non-line-of-sight (NLoS) wireless transmission. In this paper, we focus on IRS-aided radar estimation performance of a moving hidden or NLoS target. Unlike prior works that employ a single IRS, we investigate this problem using multiple IRS platforms and assess the estimation performance by deriving the associated Cram\'er-Rao lower bound (CRLB). We then design Doppler-aware IRS phase shifts by minimizing the scalar A-optimality measure of the joint parameter CRLB matrix. The resulting optimization problem is non-convex, and is thus tackled via an alternating optimization framework. Numerical results demonstrate that the deployment of multiple IRS platforms with our proposed optimized phase shifts leads to  a higher estimation accuracy compared to  non-IRS and single-IRS alternatives.
\end{abstract}

\begin{IEEEkeywords}
A-optimality, hidden target sensing, intelligent reflecting surfaces, parameter estimation, radar.
\end{IEEEkeywords}
\IEEEpeerreviewmaketitle
\section{Introduction}
\IEEEPARstart{}{}  
In recent years, intelligent reflecting surfaces (IRS) have emerged as a promising technology for smart wireless environments \cite{renzo2019smart,hodge2020intelligent}. An  IRS consists of low-cost passive meta-material elements capable of  varying the phase of the impinging  signal and  hence shaping the radiation beampattern to alter the radio propagation environment. Initial research on IRS was limited to wireless communication applications such as range extension to users with obstructed direct links \cite{wu2019towards}, joint wireless information and power transmission %in internet of things (IoT) networks 
\cite{kumar2022feasibility}, physical layer security~\cite{mishra2022optm3sec}, unmanned air vehicle (UAV) communications~\cite{Giovanni2021}, and shaping the wireless channel through multi-beam design %in millimeter-wave communications 
\cite{torkzaban2021shaping}. % see e.g.~\cite{wu2019towards,kumar2022feasibility,torkzaban2021shaping} and the references therein. 
Recent works have also introduced IRS to integrated  communications and sensing systems~\cite{wang2022stars,wei2022irs,mishra2022optm3sec,elbir2022rise}. In this paper, we focus on IRS-aided sensing, following  the advances made in~\cite{esmaeilbeig2022irs,wang2022stars}.

The literature on IRS-aided radar \cite{aubry2021reconfigurable,wei2022irs} is primarily  focused on the radar's ability to sense objects that are hidden from its line-of-sight (LoS). 
%The goal of a radar system is to acquire (or preserve) the maximum amount of information from the desirable sources in the environment; where in  fact, the transmit signal can be viewed as a  medium that collects information. The target detection and estimation performance  of the radar system is shown to be improved by a judicious design of the  probing signals and processing schemes. In the context of  IRS-aided radar, the optimization of the IRS reflections can considerably improve  the radar performance.
While there is a rich body of research on non-IRS-based non-line-of-sight (NLoS) radars (see, e.g.,~\cite{watson2019non} and the references therein), the proposed  formulations  require prior and  rather accurate knowledge of the geometry of propagation environment. In contrast, IRS-aided radar utilizes the signals received from the %radar-IRS-target-IRS-radar or  
NLoS  paths to compensate for the end-to-end transmitter-receiver or LoS path loss \cite{esmaeilbeig2022irs}. 

The potential of IRS in enhancing the estimation performance of radar systems has been recently investigated in~\cite{song2022intelligent,wang2022stars,wei2022multi}.
Some recent studies such as~\cite{song2022intelligent}, employ IRS to correctly estimate the direction-of-arrival (DoA) of a stationary target. 
Nearly all of the aforementioned works consider a single-IRS aiding the  radar for estimating the parameters of a stationary target. The IRS-based radar-communications in \cite{wei2022multi} included moving targets but did not examine parameter estimation. In this paper, we focus on the estimation performance of a multi-IRS-aided radar dealing with moving targets. 

In particular, we jointly estimate target reflectivity and Doppler velocity with multiple IRS platforms in contrast to the scalar parameter estimation via a single IRS in~\cite{song2022intelligent}. We derive the Cram\'er-Rao lower bound (CRLB) for these parameter estimates and then determine the optimal IRS phase shifts using CRLB as a benchmark. %While the IRS phase-shifts designed in~\cite{song2022intelligent} are based on  the  CRLB of DoA, in this work  we  design Doppler-resilient phase-shifts for  IRS based on CRLB of Doppler estimation. 
Previously, maximization of signal-to-noise ratio (SNR) or signal-to-interference-to-noise ratio (SINR) was employed in \cite{buzzi2022foundations} to determine the optimal phase shifts for target detection. However, optimization of the SNR or SINR does not guarantee  an improvement in  target estimation accuracy. Our previous works~\cite{esmaeilbeig2022irs} and~\cite{esmaeilbeig2022joint}  introduced multi-IRS-aided radar for NLoS sensing of a stationary target and  derived the best linear unbiased estimator (BLUE) focusing only on target reflectivity and the  CRLB of  DoA, respectively. 

The rest of the paper is organized as follows. In the next section, we introduce the signal model for the multi-IRS-aided radar. In section~\ref{sec_2}, we derive the CRLB for joint parameter estimation. section~\ref{sec_4} presents the algorithm to optimize the IRS phase shifts. We evaluate our methods via numerical experiments in section~\ref{sec_5} and conclude the paper in section~\ref{sec:conclusion}.

Throughout this paper, we use bold lowercase and bold uppercase letters for vectors and matrices, respectively. $\mathbb{C}$ and $\mathbb{R}$ represent the set of complex and real numbers, respectively. $(\cdot)^{\top}$ and $(\cdot)^{\mathrm{H}}$ denote the vector/matrix transpose, and the Hermitian transpose, respectively. The trace of a matrix is denoted by $\operatorname{Tr}(.)$. $\textrm{Diag}(.)$ denotes the diagonalization operator that produces a diagonal matrix  with same diagonal entries as the entries of its  vector argument, while $\textrm{diag}(.)$ outputs a vector containing the  diagonal entries of the input matrix. The $mn$-th element of the matrix $\mbB$ is $\left[\mbB\right]_{mn}$. The Hadamard (element-wise) and Kronecker products are denoted by notations $\odot$ and $\otimes$, respectively. 
%The vectorized form of a matrix $\mbB$ is written as $\operatorname{vec}(\mbB)$.
The element-wise matrix derivation operator is  $[\frac{\partial\mbA}{\partial b}]_{ij}=\frac{\partial\mbA_{ij}}{\partial b}$. 
The $s$-dimensional all-one vector and the identity matrix of  size $s\times s$ are denoted as $\mathbf{1}_{s}$ and $\mbI_s$, respectively. Finally, $\Re{\cdot}$ and $\Im{\cdot}$ return the %phase arguments, 
the real part, and the imaginary part of a complex input vector, respectively.
%--------------------------------------------------
\begin{comment}
\begin{table}[t]
\caption{Comparison with the state-of-the-art}
\begin{center}
\begin{threeparttable}
\begin{tabular}{c c c c}
\toprule
cf.& IRS & Target & Design metric\\ 
 \hline
 \cite{liu2022joint} & Single& Stationary, single&SINR\\
 \hline
 \cite{wei2022multi} & Double& Moving, single&SINR\\
 \hline
\cite{song2022intelligent} & Single& Stationary, single&SINR and CRLB\\
 \hline
 \cite{wang2022stars} & Single& Stationary, single&CRLB\\
 \hline
This paper & Multiple & Moving, single&CRLB\\ 
 \hline
\end{tabular}
\end{threeparttable}
\end{center}
\end{table}
\end{comment}
%--------------------------------------------------
\nocite{eamaz2021modified}\vspace{-10pt}
\section{System Model}
\label{sec_1}
%\textcolor{red}{why the title of the section about `stationary targets'. Isn't the paper about moving targets??}
Consider a single-antenna pulse-Doppler IRS-aided radar, %consisting of collocated transmitter and receiver and 
which transmits a train of $N$ uniformly-spaced pulses $s(t)$, \ze{each of which is nonzero over the support $[0, \tau]$}, with  the pulse repetition interval (PRI) $T_{_{p}}$; its reciprocal $1/T_{_{p}}$ is  the pulse repetition frequency (PRF). The  transmit signal is
\par\noindent\small
\begin{equation}
x(t)=\sum_{n=0}^{N-1}s(t-nT_{_{p}}), \;   0\leq t \leq (N-1)T_{_{p}}.
\end{equation}\normalsize
The entire duration of all $N$ pulses is the coherent processing interval (CPI) following a slow-time coding procedure~\cite{mishra2019sub}. 

\begin{figure}[t]
	\centering
	%\vspace{-0.25in}
    \includegraphics[width=0.62\columnwidth]{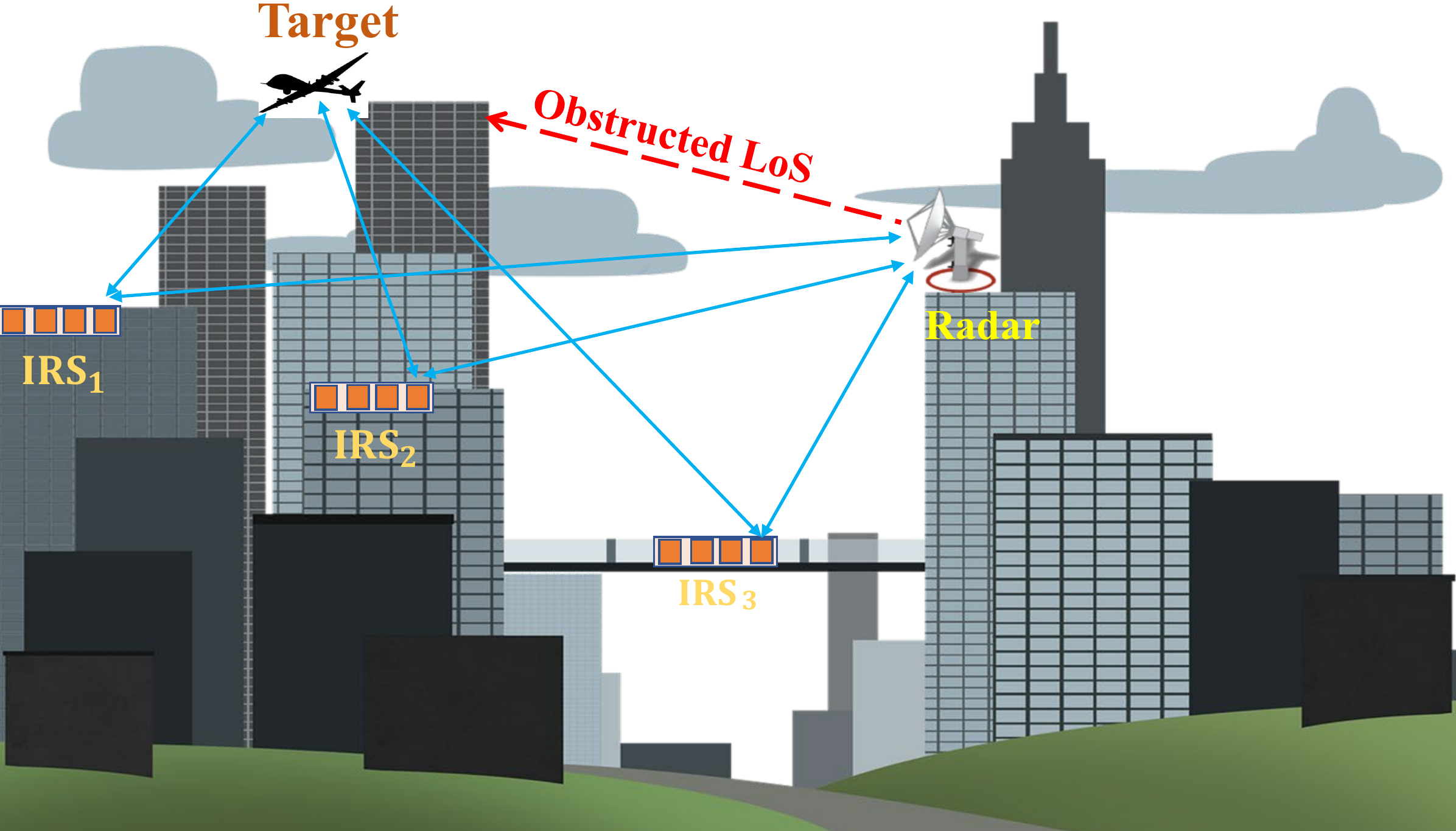}
	\caption{\textcolor{black}{A simplified illustration of various NLoS or virtual LoS links between the radar and the hidden, moving target, provided by multiple IRS platforms mounted on urban infrastructure.}
	}
	%\Ae{Completely pointless!}
	\label{fig::4}
\end{figure}

Assume that the propagation environment has $K$ IRS platforms, each with $M$ reflecting elements, deployed on stationary platforms at known locations (Fig.~\ref{fig::4}). Each of the $M$ reflecting elements in the $k$-th IRS or IRS$_k$ reflects the incident signal with a phase shift and amplitude change that is configured via a smart controller~\cite{bjornson2022reconfigurable}. Denote the phase shift matrix of IRS$_k$ by\par\noindent\small
\begin{equation}
\label{eq:4}
\bPhi_k=\text{\textrm{Diag}}\left(\beta_{k,1}e^{\textrm{j}\phi_{k,1}},\ldots,\beta_{k,M}e^{\textrm{j}\phi_{k,M}}\right),
\end{equation}\normalsize
where $\phi_{k,m}\in[0,2\pi]$, and $\beta_{k,m}\in [0,1]$ are, respectively, the phase shift and  the amplitude reflection gain associated with the $m$-th passive element of  IRS$_k$. In practical settings, it usually  suffices to design only the phase shifts so that  $\beta_{k,m}=1$ for all $(k,m)$.  
The radar-IRS$_k$-target-IRS$_k$-radar channel coefficient/gain is \par\noindent\small
\begin{equation}
\label{eq:CSI-vec}
	h_{_{\NLoS,k}}= \mbb^{\T}(\theta_{ir,k}) \bPhi_k \mbb(\theta_{ti,k}) \mbb^{\T}(\theta_{ti,k}) \bPhi_k \mbb(\theta_{ir,k}),
\end{equation}\normalsize
where $\theta_{ir,k}$ ($\theta_{ti,k}$) is the angle between the radar (target) and IRS$_k$, and  each IRS is a uniform linear array with the inter-element spacing $d$, %and the same number of reflecting elements $M$,
with the steering vector \par\noindent\small
\begin{align}\label{eq:str}
\mbb(\theta)&= \left[1,e^{\mathrm{j}\frac{2\pi d}{\lambda} \sin \theta},\ldots,e^{\mathrm{j}\frac{2\pi d}{\lambda}(M-1) \sin \theta}\right]^{\top},
\end{align}\normalsize
and $\lambda$ denoting the carrier wavelength. 

Consider a single Swerling-0 model \cite{skolnik2008radar}, moving target  with $f_{_{D_0}}$ and $\alpha_0$ being, respectively, its Doppler frequency and target back-scattering coefficient as observed via the LoS path between the target and radar. Denote the same parameters by, respectively, $f_{_{D_k}}$ and $\alpha_{_k}$, for $k \in \left\{1,\ldots,K\right\}$ as observed by the radar from the NLoS path via IRS$_k$. All Doppler  frequencies lie in the unambiguous frequency  region, i.e., %the Doppler  frequencies are 
up to  PRF. The received signal in a known range-bin is a superposition of echoes from the LoS and NLoS paths as \par\noindent\small
\begin{align}
\label{eq:cont_rx}
 y(t)&=\alpha_{0}h_{_{\LoS}} \sum_{n=0}^{N-1}s(t-nT_{_{p}}) e^{\mathrm{j} 2\pi f_{D_0} t}\nonumber\\
 &+\sum_{k=1}^{K}\sum_{n=0}^{N-1}\alpha_{k} h_{_{\NLoS,k}}s(t-nT_{_{p}}) e^{\mathrm{j}2\pi f_{_{D_k}} t} +w(t)\nonumber\\
 &\approx
 \alpha_{0}h_{_{\LoS}} \sum_{n=0}^{N-1}s(t-nT_{_{p}}) e^{\mathrm{j} 2\pi f_{_{D_0}}n T_p }\nonumber\\
 &+\sum_{k=1}^{K}\sum_{n=0}^{N-1}\alpha_{k} h_{_{\NLoS,k}}s(t-nT_{_{p}}) e^{\mathrm{j}2\pi f_{_{D_k}}n T_p } +w(t)
\end{align}\normalsize 
where $h_{_{\LoS}}$ is the  radar-target-radar LoS channel state information (CSI), $h_{_{\NLoS,k}}$ is the NLoS channel through  radar-IRS$_k$-target-IRS$_k$-radar path, $w(t)$ is the random additive signal-independent interference, and the last approximation %in~\eqref{eq:cont_rx} we 
assumes  the target is moving with low speed $f_{_{D_k}} \ll 1/\tau$ for $k \in \left\{0,\ldots,K\right\}$ and  have low acceleration, so that the phase rotation within the CPI could be approximated as a constant.\color{black}

\ze{
We design a radar system to  sense a moving  target  in the two-dimensional (2-D) Cartesian plane. Our proposed  signal model and methods are readily extendable to 3-D scenarios by replacing the 1-D DoA with 2-D DoA in ~\eqref{eq:CSI-vec} and~\eqref{eq:str}.} In particular, we consider a  target tracking scenario where the radar wishes to examine a range-cell for a potential target~\cite{liu2021cramer,naghsh2014radar}. 
Accordingly, \textcolor{black}{we collect $N$ slow-time samples of \eqref{eq:cont_rx} corresponding to $N$ pulses} in the vector  \par\noindent\small
 \begin{equation}
 \label{eq:2}
\mby=\alpha_{0}h_{_{\LoS}}\left(  \mbx  \odot \mbp(\nu_0) \right)+\sum_{k=1}^{K}\alpha_{_{k}}h_{_{\NLoS,k}}\left( \mbx  \odot \mbp(\nu_k)\right) +\mbw,
\end{equation}\normalsize
where $\nu_k= 2 \pi f_{_{D_k}} T_p\in [-0.5,0.5)$ for  $k \in \{0,\ldots,K\}$ are the normalized Doppler  shifts, $\mbp(\nu)=\left[1,e^{\mathrm{j}\nu} ,\ldots,e^{\mathrm{j}(N-1)\nu  }\right]^{\top}$
 is a generic  Doppler steering vector, and $\mbx=[x(0),x(T_{_{p}}),\ldots,x((N-1)T_{_{p}})]^{\T}$ and $\mbw =[w(0),w(T_{_{p}}),\ldots,w((N-1)T_{_{p}})]^{\T}$ are the $N$-dimensional column vectors of the transmit signal and  the zero-mean random noise vector with a  covariance $\mbR$, respectively~\cite{vargas2022dual}. 
 
\ze{%Assume the  complex target back-scatter coefficient  of the target observed from the LoS and NLoS paths are  collected in  the vector $\balpha=[\alpha_{_{0}}, \alpha_{_{1}},\alpha_{_{2}},\ldots,\alpha_{_{K}}]^{\top}$ and the corresponding channel coefficients are  collected in $\mbh=[h_{_{\LoS}},h_{_{\NLoS,1}},\ldots,h_{_{\NLoS,K}}]^{\T}$.
 Assume the complex target back-scattering coefficients observed from the LoS and NLoS paths are collected in  the vector $\balpha=[\alpha_{_{0}}, \alpha_{_{1}},\alpha_{_{2}},\ldots,\alpha_{_{K}}]^{\top}$.  The corresponding channel coefficients are collected in 
 $\mbh=[h_{_{\LoS}},h_{_{\NLoS,1}},\ldots,h_{_{\NLoS,K}}]^{\T}$.} Denote the  sensing matrix by $\mbA=\left[\mba_0, \mba_1,\ldots,\mba_K\right]\in \mathbb{C}^{N \times K+1}$, where $\mba_0 =h_{_{\LoS}} \left[ \mbx \odot \mbp(\nu_0) \right]$ and $\mba_k =h_{_{\NLoS,k}} \left[ \mbx \odot \mbp(\nu_k) \right]$. This implies that $\mbA=\mbx \mbh^{\T} \odot \mbP(\nu)$, where the Doppler shift matrix is $\mbP(\bnu) = [\mbp(\nu_0),\mbp(\nu_1),\ldots,\mbp(\nu_K)]\in \mathbb{C}^{N \times K+1}$. Then, the received signal \eqref{eq:2} in the vector form is given as\par\noindent\small
\begin{equation}
\label{eq:SISO}	
\mby=\mbA \balpha +\mbw,
\end{equation}\normalsize
 
In presence of multiple-IRS platforms (Fig.\ref{fig::4}), we  define the 
\emph{LoS-to-NLoS link strength ratio  (LSR)}  as %\par\noindent\small
%\begin{equation}\label{eq::gamma}
	%\gamma=\frac{|\alpha_{_0} h_{\LoS}|^2}{\|\balpha^{\T} \mbh_{_\NLoS}\|^2_{_2}},
	$\gamma=\frac{|\alpha_{_0} h_{_{\LoS}}|^2}{\sum_{k=1}^{K}|\alpha_k h_{_{\NLoS,k}}|^2}$, 
%\end{equation}\normalsize
which governs the relative strengths of the signals received from both paths. In cases where the LoS path is obstructed or weaker than the NLoS paths as  illustrated in Fig.~\ref{fig::4}, we have  $h_{_{\LoS}} \ll h_{_{\NLoS,k}}$.
%--------------------------------------------------
\section{CRLB for Multi-IRS-Aided Radar}%aided radar}
\label{sec_2}
 %We analyze the Cram\'er-Rao lower bound (CRLB) of the IRS-aided target parameter estimation problem.
 
 %For an unbiased estimator of the vector parameter $\bzeta$, the covariance matrix of  $\hat{\bzeta}$ is lower bounded as $\bC_{\hat{\bzeta}}=\mathbb{E}\left\{\left(\hat{\bzeta}-\bzeta\right)\left(\hat{\bzeta}-\bzeta\right)^{\H}\right\}\geq \bC_{_{\textrm{CRLB}}}$,
 %in the sense that the difference  $\bC_{\hat{\bzeta}}-\bC_{_{\textrm{CRLB}}}$ is a positive semidefinite matrix \cite{bickel2015mathematical}. 
 Denote the $3(K+1) \times 1$ vector of unknown target parameters as $\bzeta=\left[\Tilde{\balpha}^{\T}, \bnu^{\T}\right]^{\T}$,
 where $\Tilde{\balpha}=[\balpha_{R}^{\top},\balpha_I^{\top}]^{\T}$, with $\balpha_{R}=\Re{\balpha}$, $\balpha_I=\Im{\balpha}$, and $\bnu=[\nu_0,\nu_1,\ldots,\nu_K]^{\T}$ denoting the Doppler shifts vector. 
 For an unbiased estimator  $\hat{\bzeta}$, the covariance matrix is lower bounded as $\bC_{\hat{\bzeta}}=\mbE\{{(\hat{\bzeta}-\bzeta)(\hat{\bzeta}-\bzeta)^H}\}\geq \bC_{_{\textrm{CRLB}}}$, in the sense that the difference  $\bC_{\hat{\bzeta}}-\bC_{_{\textrm{CRB}}}$ is a positive semidefinite matrix. 
 %To derive the CRLB of $\bzeta$, we first formulate the FIM for all the corresponding unknown parameters in $\bzeta$. We divide the FIM and $\mbC_{_{\textrm{CRLB}}}$ into submatrices as
Express the $3(K+1) \times 3(K+1)$ Fisher information matrix (FIM) in terms of the parameter submatrices as\par\noindent\small
\begin{equation}
\label{eq:fisher-submatrices}
		\mbF=\begin{bmatrix}
			\mbF_{\Tilde{\balpha},\Tilde{\balpha}}& \mbF_{\Tilde{\balpha},\bnu}\\
			\mbF_{\bnu,\Tilde{\balpha}}&\mbF_{\bnu,\bnu}
		\end{bmatrix}=\begin{bmatrix}
			\bC_{\Tilde{\balpha},\Tilde{\balpha}}& \bC_{\Tilde{\balpha},\bnu}\\
			\bC_{\bnu,\Tilde{\balpha}}&\bC_{\bnu,\bnu}
		\end{bmatrix}^{-1}=\bC_{_{\textrm{CRLB}}}^{-1},
\end{equation}\normalsize
The following Theorem~\ref{lem:submat} derives each submatrix term.
\begin{theo}\label{lem:submat}
\ze{For the  signal model in~\eqref{eq:SISO}, where $\mbw \sim \mathcal{CN}\left(\bzero,\mbR\right) $}, the submatrices of the FIM in~\eqref{eq:fisher-submatrices} are \par\noindent\small
\begin{align}
&\mbF_{\Tilde{\balpha},\Tilde{\balpha}}=2\Re{\left[1 \; \mathrm{j}\right]^\H\otimes \left[1 \; \mathrm{j}\right]\otimes \left(\mbA^{\mathrm{H}}\mbR^{-1} \mbA\right)},\;\in\mathbb{R}^{2K \times 2K}, \label{eq:submatrices-tilde-alpha}
\end{align}
\begin{align}
&\mbF_{\Tilde{\balpha},\bnu}= \mbF^{\T}_{\bnu,\Tilde{\balpha}}\nonumber\\
&=2\Re{\left (\left[1 \; \mathrm{j}\right]^{\mathrm{H}} \otimes\left(\mbA^{\mathrm{H}} \mbR^{-1} \dot\mbA\right)\right )\left(\balpha \otimes \mbI_{K}\right)},\;\in\mathbb{R}^{2K \times K}, \label{eq:submatrices-tilde-alpha-nu}\\
&\mbF_{\bnu,\bnu}=2\Re{\left(\balpha \otimes \mbI_{K}\right)^{\mathrm{H}}\left(\dot \mbA^{\mathrm{H}} \mbR^{-1} \dot \mbA\right)\left(\balpha \otimes \mbI_{K}\right)}\;\in\mathbb{R}^{K \times K}.\label{eq:submatrices-nu-nu}
\end{align}
\end{theo}\normalsize
\begin{IEEEproof}Using the Slepian-Bangs formula \cites{bangs1971array}[p. 360]{stoica2005spectral}[p. 565]{kay1993fundamentals}, for the complex observation vector $\mby \sim \mathcal{CN}\left(\bmu,\mbR\right)$, the $(m,n)$-th element of the FIM $\mbF$ is\par\noindent\small
\begin{equation}
\label{eq:slepian}
[\mbF]_{mn}=\operatorname{Tr} \left(\mbR^{-1}\frac{\partial \mbR }{\partial\bzeta_m} \mbR^{-1}\frac{\partial \mbR }{\partial\bzeta_n}\right)+
2\Re{\frac{\partial\bmu}{\partial\bzeta_m}^H\mbR^{-1}\frac{\partial\bmu}{\partial\bzeta_n}}.
\end{equation}\normalsize
From \eqref{eq:SISO}, we have $\bmu = \mbA \balpha$~\cite{kay1993fundamentals}. Define \par\noindent\small
\begin{align}\label{eq::ai}
		\frac{\partial\bmu}{\partial\balpha_{R_k}}=\frac{\partial (\mbA \balpha)}{\partial\alpha_{R_k}}=\mba_k \text{ and }%\nonumber \\
		\frac{\partial\bmu}{\partial\balpha_{I_k}}=\frac{\partial (\mbA \balpha)}{\partial\balpha_{I_k}}=\mathrm{j}\mba_k.
\end{align}\normalsize
Denote
$\dot \mbA=\left [\frac{\partial \mbA}{\partial\nu_0},\ldots,\frac{\partial \mbA}{\partial\nu_k}\right ]$ and $\dot \mbP(\bnu)=\sum_{k=0}^{K}\frac{\partial\mbP(\bnu)}{\partial\nu_k}=[\frac{\partial\mbp(\nu_0)}{\partial\nu_0},\ldots,\frac{\partial\mbp(\nu_K)}{\partial\nu_K}]$. It follows that\par\noindent\small
\begin{align}
	\label{eq::nu}
		\frac{\partial\bmu}{\partial\nu_k}&=\left [\mbx \mbh^{\T} \odot \dot\mbP(\bnu) \right ] \be_k \be_k^{\top} \balpha =\dot \mbA \left [\be_k \otimes \balpha    \right],
\end{align}\normalsize
where $\be_k$ is a $(K+1) \times 1$ vector whose $k$-th element is unity and remaining elements that are zero. Rewrite $\mbF_{\Tilde{\balpha},\Tilde{\balpha}}$ as\par\noindent\small
\begin{equation}
\label{eq::fisher-submatrices_alpha}
\mbF_{\Tilde{\balpha},\Tilde{\balpha}}=\begin{bmatrix}
			\mbF_{\balpha_R,\balpha_R}& \mbF_{\balpha_R,\balpha_I}\\
			\mbF_{\balpha_I,\balpha_R}&\mbF_{\balpha_I,\balpha_I}
\end{bmatrix}.
\end{equation}\normalsize
\ze{Since the covariance matrix does not depend on the parameters to be estimated, the first term in~\eqref{eq:slepian} is zero}. We have\par\noindent\small
\begin{equation}
\begin{split}
\left[\mbF_{\balpha_{R},\balpha_{R}}\right]_{mn}&=2\Re{\frac{\partial\bmu}{\partial\balpha_{R_m}}^\H\mbR^{-1}\frac{\partial\bmu}{\partial\balpha_{R_n}}}\\
&=2\Re{\mba_m^\H \mbR^{-1} \mba_n}=2\Re{\be_m^{\top} \mbA^\H \mbR^{-1} \mbA \be_n},   
\end{split}
\end{equation}\normalsize
which yields %\par\noindent\small
%\begin{equation}
		$\mbF_{\balpha_{R},\balpha_{R}}=2\Re{ \mbA^H \mbR^{-1} \mbA}$. 
%\end{equation}\normalsize
Similarly, %\par\noindent\small
	%\begin{equation}
		$[\mbF_{\balpha_{R},\balpha_{I}}]_{mn}=2\Re{\mba_m^\H \mbR^{-1} \mathrm{j}\mba_n}=-2\Im{\be_m^\T \mbA^\H \mbR^{-1} \mbA \be_n}$.
	%\end{equation}\normalsize
%In a similar manner we get all the submatrices in~\eqref{eq::fisher-submatrices_alpha} as 
This results in\par\noindent\small
\begin{align}\label{eq::J-alpha}
\mbF_{\balpha_{I},\balpha_{I}}&=\mbF_{\balpha_{R},\balpha_{R}}=2\Re{\mbA^H \mbR^{-1} \mbA},\nonumber\\
\mbF_{\balpha_{R},\balpha_{I}}&=-\mbF_{\balpha_{I},\balpha_{R}}=-2\Im{\mbA^H \mbR^{-1} \mbA}.
\end{align}\normalsize
Substituting~\eqref{eq::J-alpha} in~\eqref{eq::fisher-submatrices_alpha} produces \eqref{eq:submatrices-tilde-alpha}. 

%Next, %we compute other submatrices in~\eqref{eq:fisher-submatrices}. 
%we have\par\noindent\small
%\begin{equation}
%\label{eq:fisher-submatrices_alpha-v}
%\mbF_{\Tilde{\balpha},\mbv}=\begin{bmatrix}
%\mbF_{\balpha_R,\nu}\\
%\mbF_{\balpha_I,\nu}
%\end{bmatrix}.
%\end{equation}\normalsize
Similarly, substituting~\eqref{eq::ai} and~\eqref{eq::nu} in~\eqref{eq:slepian} yields\par\noindent\small
\begin{align}\label{eq::alpha-nu}
[\mbF_{\balpha_{R},\bnu}]_{mn}&=2\Re{\be_m^{\top} \mbA^\H\mbR^{-1} \dot \mbA \left[\mathbf{I}_K \otimes \be_n\right] \balpha}, \nonumber\\
[\mbF_{\balpha_{I},\bnu}]_{mn}&=2\Re{-\mathrm{j}\be_m^{\top} \mbA^\H\mbR^{-1} \dot \mbA \left[\mathbf{I}_K \otimes \be_n\right] \balpha},
\end{align}\normalsize
which by considering~\eqref{eq::alpha-nu} in $\mbF_{\Tilde{\balpha},\mbv}=[\mbF_{\balpha_R,\nu}, \mbF_{\balpha_I,\nu}]^T$  %~\eqref{eq:fisher-submatrices_alpha-v}, 
yields~\eqref{eq:submatrices-tilde-alpha-nu}.

Finally, from~\eqref{eq::nu}, we obtain\par\noindent\small
\begin{equation}
\label{eq:fisher-nu-nu}
[\mbF_{\bnu,\bnu}]_{mn}= 2\Re{\left[\be_m \otimes \balpha \right]^{\H} \dot \mbA^{\H} \mbR^{-1} \dot \mbA \left[\be_n \otimes \balpha\right] }.
\end{equation}\normalsize
%Collecting  the elements of \eqref{eq:fisher-nu-nu}  into  ~\eqref{eq:submatrices-nu-nu} 
Stacking  all $(K+1)^2$ elements in~\eqref{eq:fisher-nu-nu}, we obtain \eqref{eq:submatrices-nu-nu}. %This completes the proof. 
\end{IEEEproof}
Note that, in the absence of IRS, \ze{assuming $\mbR=\sigma^2\mathbf{I}_N$}, the FIM corresponding to the LoS received signal in~\eqref{eq:2} becomes 
%\begin{align}
$\mbF_{\Tilde{\alpha_0},\Tilde{\alpha_0}}=\frac{2|h_{_{\LoS}}|^2  \|\mbx \odot \mbp(\nu_0)\|^2 }{\sigma^2}  \mathbf{I}_2$, and %\\
$\mbF_{\nu_0,\nu_0} = \frac{2|\alpha_0 h_{_{\LoS}}|^2  \|\mbx \odot \dot \mbp(\nu_0)\|^2 }{\sigma^2}$,  %\label{eq:crlb_LoS}
%\end{align}
where $\Tilde{\alpha_0}=[\Re{\alpha_0},\Im{\alpha_0}]^{\T}$ and $\dot \mbp(\nu_0)=\frac{\partial\mbp(\nu_0)}{\partial\nu_0}$. %One can observe from ~\eqref{eq:crlb_LoS} that 
The CRLB of the Doppler shift depends on the variable itself. A proper Doppler estimator based on the  maximum likelihood criterion is available in~\cite{bamler1991doppler}.\nocite{eamaz2022one,esmaeilbeig2020deep} %, that can be  employed. %In the  following, we  discuss a  CRLB optimization  approach to  Doppler-aware IRS phase-shifts design.
\section{Doppler-Aware Phase Shift Design} %under A-optimality}%  design for error bound minimization}
\label{sec_4}
 %We now determine the phase-shifts of each IRS %multi-IRS aided radar system, we determine the IRS parameters $\bPhi_k$, $k \in \{1,\ldots,K\}$ 
 %by minimizing the lower error bound. 
 Since we are estimating a vector of parameters, the CRLB is a matrix. In such a situation, to obtain a scalar objective function, % need a scalar measure of the  CRLB to be  used as the objective of optimization. In 
 \cite{tohidi2018sparse} introduced several CRLB-based scalar metrics for optimization. In this paper, we choose the objective function as $\operatorname{Tr}\left(\bC_{_{\textrm{CRLB}}}\right)$ based on an average sense or the A-optimality criterion.
 %Note that the trace of a $\bC_{_{\textrm{CRLB}}}$is the sum of all scalar  CRLBs presented in~\eqref{eq:schur1} and ~\eqref{eq:schur2}.
 The A-optimality criterion has a low computational complexity for optimization\cite{tohidi2018sparse} and directly  targets the minimization of  the parameters of interest. % in comparison with  the other  metrics introduced in~\cite{tohidi2018sparse}, such as  $\lambda_{max}(\bC_{_{\textrm{CRLB}}})$ and  $\textrm{det}(\bC_{_{\textrm{CRLB}}})$. 
 %The resulting optimization problem to design the phase-shifts is\par\noindent\small
 %\begin{equation}
 %\begin{aligned}
 % \label{Ne1}
 % \underset{\bPhi_k, \; k \in \{1,\ldots,K\}}{\textrm{minimize}} \quad 
%&\operatorname{Tr}\left(\bC_{_{\textrm{CRLB}}}\right)\\
%\text{subject to} \quad & \left|\left[\bPhi_k\right]_{ii}\right|=1,\;
%\quad  \left[\bPhi_k\right]_{ij}=0, i\neq j.
% \end{aligned}
% \end{equation}\normalsize
%One can simplify the optimization problem 
%The problem in (\ref{Ne1}) according to the following suboptimality relation:
%$\operatorname{Tr}\left(\bC_{_{\textrm{CRLB}}}\right)\geq \Tr{\mbF_{\Tilde{\balpha},\Tilde{\balpha}}^{-1}} + \Tr{\mbF_{\bnu,\bnu}^{-1}}$, where
\ze{From \cite[Proposition 1]{bobrovsky1987some}, $\Tr{\mbF_{\Tilde{\balpha},\Tilde{\balpha}}^{-1}} + \Tr{\mbF_{\bnu,\bnu}^{-1}}$ is both a lower bound and an approximation of $\operatorname{Tr}\left(\bC_{_{\textrm{CRLB}}}\right)$.} In particular, we consider  the  following CRLB approximation-based optimization problem\par\noindent\small
%Since \eqref{eq:schur1} and \eqref{eq:schur2} provide  an approximation of  %$\operatorname{Tr}\left(\bC_{_{\textrm{CRLB}}}\right)$, the resulting optimization problem to design the %phase-shifts is \begin{equation}
 \begin{align}
 \label{Neg2}
\underset{\bPhi_k,\; k \in \{1,\ldots,K\}}{\textrm{minimize}} \quad  &\Tr{\mbF_{\Tilde{\balpha},\Tilde{\balpha}}^{-1}} + \Tr{\mbF_{\bnu,\bnu}^{-1}}\nonumber\\
\text{subject to} \quad & \left|\left[\bPhi_k\right]_{ii}\right|=1,\;
\quad \left[\bPhi_k\right]_{ij}=0, i\neq j.
 \end{align}
% \end{equation}
\normalsize
\ze{Our numerical  experiments validate that minimizing the objective function in~\eqref{Neg2} also minimizes $\operatorname{Tr}\left(\bC_{_{\textrm{CRLB}}}\right)$.} \ze{Note that, following the conventional radar signal processing assumptions, target  parameters do not change during the CPI~\cite{mishra2019sub}. We optimize the phase shifts for the current CPI and fixed  location of multiple IRS platforms.} The unimodularity constraint in problem \eqref{Neg2} makes it non-convex with respect to $\left\{\bPhi_k\right\}$. Moreover, the objective function of (\ref{Neg2}) is quartic with respect to $\left\{\bPhi_k\right\}$ and $\mbv_k=\diag{\bPhi_k}$. Therefore, to lay the ground for a simpler form of (\ref{Neg2}), we introduce the auxiliary variable $\mbh$  in addition to $\mbv_k=\diag{\bPhi_k}$ as follows: \par\noindent\small
\begin{align}
\label{Nega3}
\underset{\mbh,\mbv_k}{\textrm{minimize}}\quad &\Tr{\mbF_{\Tilde{\balpha},\Tilde{\balpha}}^{-1}} + \Tr{\mbF_{\bnu,\bnu}^{-1}} \nonumber\\
\text{subject to} \quad & h_k= \mbv_k^{\top} \mbS\mbv_k, \;  k \in \left\{1,\ldots,K\right\},\nonumber\\
& |[\mbv_k]_m|=1, \;  m \in \left\{1,\ldots,M\right\}, 
\end{align}
\normalsize
where $\mbS=(\mbb(\theta_{ri,k})\odot\mbb(\theta_{ti,k}))(\mbb(\theta_{ri,k})\odot\mbb(\theta_{ti,k}))^{\T}$, obtained by an  algebraic manipulation on~\eqref{eq:CSI-vec}, is a known matrix given the locations of radar, IRS platforms, and the hidden target in a 2-D plane.  
Note that by defining,
$\mbZ=[\be_1\be_1^{\top},\ldots,\be_{_{K+1}}\be_{_{K+1}}^{\top}]$,
$\mbG=\mbP(\bnu)^{\H} \Diag{\mbx}\Diag{\mbx} \mbP(\bnu)$ and
$\dot \mbG=\dot\mbP(\bnu)^{\H} \Diag{\mbx}\Diag{\mbx} \dot\mbP(\bnu)$, we rewrite blocks of the FIM defined in \eqref{eq:submatrices-tilde-alpha} and \eqref{eq:submatrices-nu-nu} based on  $\mbh$ as\par\noindent\small
\begin{align}
\mbF_{\Tilde{\balpha},\Tilde{\balpha}}&= \frac{2}{\sigma^2} \Re{\left[1 \; \mathrm{j}\right]^{\mathrm{H}}\otimes \left[1 \; \mathrm{j}\right]\otimes \left((\mbh\mbh^{\mathrm{H}})^\T \odot \mbG \right)}\label{eq:f1},\\
\mbF_{\bnu,\bnu}&=\frac{2}{\sigma^2}\Re{\left(\balpha \otimes \mbI_{K}\right)^{\mathrm{H}}\mbZ^\H \left((\mbh\mbh^{\mathrm{H}})^\T \odot \dot \mbG \right) \mbZ\left(\balpha \otimes \mbI_{K}\right)},\label{eq:f2}
\end{align}\normalsize
where %we assumed 
$\mbR=\sigma^2 \mbI_{N}$. For the unimodularity constraint in~\eqref{Nega3}, we write $\{\mbv_k\}$ as $\left\{e^{\textrm{j}\phi_{k,1}},
\ldots,e^{\textrm{j}\phi_{k,M}}\right\}$ and optimizing \eqref{Nega3} with respect to $\left\{\phi_{k,1},\ldots,\phi_{k,M}\right\}$ instead of $\mbv_{k}$. %To tackle (\ref{Nega3}), we at first, write its 
Denote the  Lagrangian function and multipliers by $f\left(\mbh\right)=\Tr{\mbF_{\Tilde{\balpha},\Tilde{\balpha}}^{-1}} + \Tr{\mbF_{\bnu,\bnu}^{-1}}$ and $\left\{\eta_{k}\right\}$. The Lagrangian formulation of (\ref{Nega3}) is %\textcolor{red}{before and after every equation use par, noindent, and small commands and, immediately after the equation, use normalsize command. This will help in saving space. See example below}
\par\noindent\small
\begin{align}
\label{eq:Ne4}
\underset{ \substack{ \mbh,\phi_{k,m},\\ k \in \{1,\ldots,K\},m \in \{1,\ldots,M\}} }{\textrm{minimize}} f\left(\mbh\right)+\sum^{K}_{k=1}\eta_{k} \left|h_{k}-\mbv_k^{\top} \mbS\mbv_k\right|^2.
\end{align}
\normalsize
The non-convex problem \eqref{eq:Ne4} may be solved via a generic nonlinear  programming solver by employing multiple random initial points 
but with increased computational complexity. % However, deploying the interior point method to solve the multivariate optimization problem such as \eqref{eq:Ne4}, can increase the computational complexity.
To avoid the high computational complexity, we resort to a task-specific \emph{alternating optimization} (AO) procedure, wherein we optimize~\eqref{eq:Ne4} over $\mathbf{h}$ and $\{\phi_{k,m}\}$ cyclically. % (i) initialize $\mbh$, and $\{\phi_{k,m}\}$, (ii) optimize \eqref{eq:Ne4} with respect to $\mbh$ in $i$th iteration as $\mbh^{(i)}$, (iii) compute $\{\phi_{k,m}^{(i)}\}$ from \eqref{eq:Ne4} at which $\mbh^{(i)}$ are used in the objective function.
%AO for \eqref{eq:Ne4} is summarized in 
Algorithm~\ref{algorithm_1} summarizes the steps of this procedure.
%-------------------------------------------
\begin{algorithm}[H]
	\caption{IRS phase shift design via A-optimality. }
    \label{algorithm_1}
    \begin{algorithmic}[1]
    \Statex \textbf{Input:} Initialization values $\mbh^{(0)}$, and $\left\{\phi_{k,m}^{(0)}\right\}$, the total number of iterations $\mathcal{E}$, and $\varepsilon$ as a small positive number.
    \Statex \textbf{Output:} Optimized phase shifts $\bPhi^*_k$, $k \in \{1,\ldots,K\}$. %\textcolor{red}{you still need to mention the output variables here but they should exactly match the returned variables}
    \State Set $\mbh^{(i)}$, and $\left\{\phi_{k,m}^{(i)}\right\}$, the parameters at the $i$th iteration.
      \State Evaluate $\mbF_{\Tilde{\balpha},\Tilde{\balpha}}^{-1}$ and $\mbF_{\bnu,\bnu}^{-1}$ as in (\ref{eq:f1}) and (\ref{eq:f2}).
      
       \State %Evaluate $\left\{\mbv_{k}\right\}$, based on 
       $\mbv_{k} \leftarrow \diag{\bPhi_k}$, where %$\bPhi_k$ is \par\noindent\small
     %\begin{equation}
%\label{eq:zand}
	$\bPhi_k=\text{\textrm{Diag}}\left(e^{\textrm{j}\phi_{k,1}},\ldots,e^{\textrm{j}\phi_{k,M}}\right)$.
%\end{equation}\normalsize
     \State Set the objective function as\par\noindent\small
     \begin{equation}%\label{eq:obj_main}
       g\left(\mbh,\left\{\phi_{k,m}\right\}\right)\leftarrow  \Tr{\mbF_{\Tilde{\balpha},\Tilde{\balpha}}^{-1}} + \Tr{\mbF_{\bnu,\bnu}^{-1}}+\sum^{K}_{k=1}\eta_{k} \left|h_{k}-\mbv_k^{\top} \mbS\mbv_k\right|^2\nonumber
     \end{equation}\normalsize
    
    \For{$i=0:\mathcal{E}-1$}
    \State \;\; %Compute $\mbh^{(i+1)}$ by  solving \par\noindent\small

$\mbh^{(i+1)} \leftarrow \underset{\mbh}{\textrm{argmin}}\quad  g\left(\mbh,\left\{\phi^{(i)}_{k,m}\right\}\right)$.

\State \;\;
$\left\{\phi^{(i+1)}_{k,m}\right\}\leftarrow\underset{\left\{\phi_{k,m}\right\}}{\textrm{argmin}}\quad  g\left(\mbh^{(i+1)},\left\{\phi_{k,m}\right\}\right)$.

\EndFor
    \State Choose Lagrangian multipliers $\left\{\eta_{k}\right\}$ such that $g\left(\mbh^{(\mathcal{E})},\left\{\phi^{(\mathcal{E})}_{k,m}\right\}\right)\leq \varepsilon$.
    \State \Return  %Approximated 
      %Optimized phase-shift matrix  
      $\bPhi^*_k \leftarrow \text{\textrm{Diag}}\left(e^{\textrm{j}\phi^{(\mathcal{E})}_{k,1}},\ldots,e^{\textrm{j}\phi^{(\mathcal{E})}_{k,M}}\right)$.
      %$\left\{\phi^{(\mathcal{E})}_{k,m}\right\}$. %from the CRLB estimation, 
    %\textcolor{red}{I do not see in the steps below where do you obtain these outputs. I only see you get $\mbh^{(\mathcal{E})}$ and $\left\{\phi^{(\mathcal{E})}_{k,m}\right\}$. Not the same! }

    \end{algorithmic}
\end{algorithm}
%----------------------------------
We used a nonlinear programming  solver based on interior-point method \cite{bazaraa2013nonlinear,bezdek2003convergence} to solve the subproblems in Steps 6 and 7. Since the objective function in steps  6 and 7 decreases monotonically, the convergence of  AO is guaranteed. 
The number of constraints in (\ref{Nega3}) increases with the number of IRS platforms, making it more likely to capture the desired local optima inside the feasible set with the objective value meeting the threshold i.e. $g\left(\mbh^{(\mathcal{E})},\left\{\phi^{(\mathcal{E})}_{k,m}\right\}\right)\leq \varepsilon$. 
\section{Numerical Experiments}
\label{sec_5}
%---------------------------------
\begin{figure}[t]
\centering
	\includegraphics[width=0.8\columnwidth]{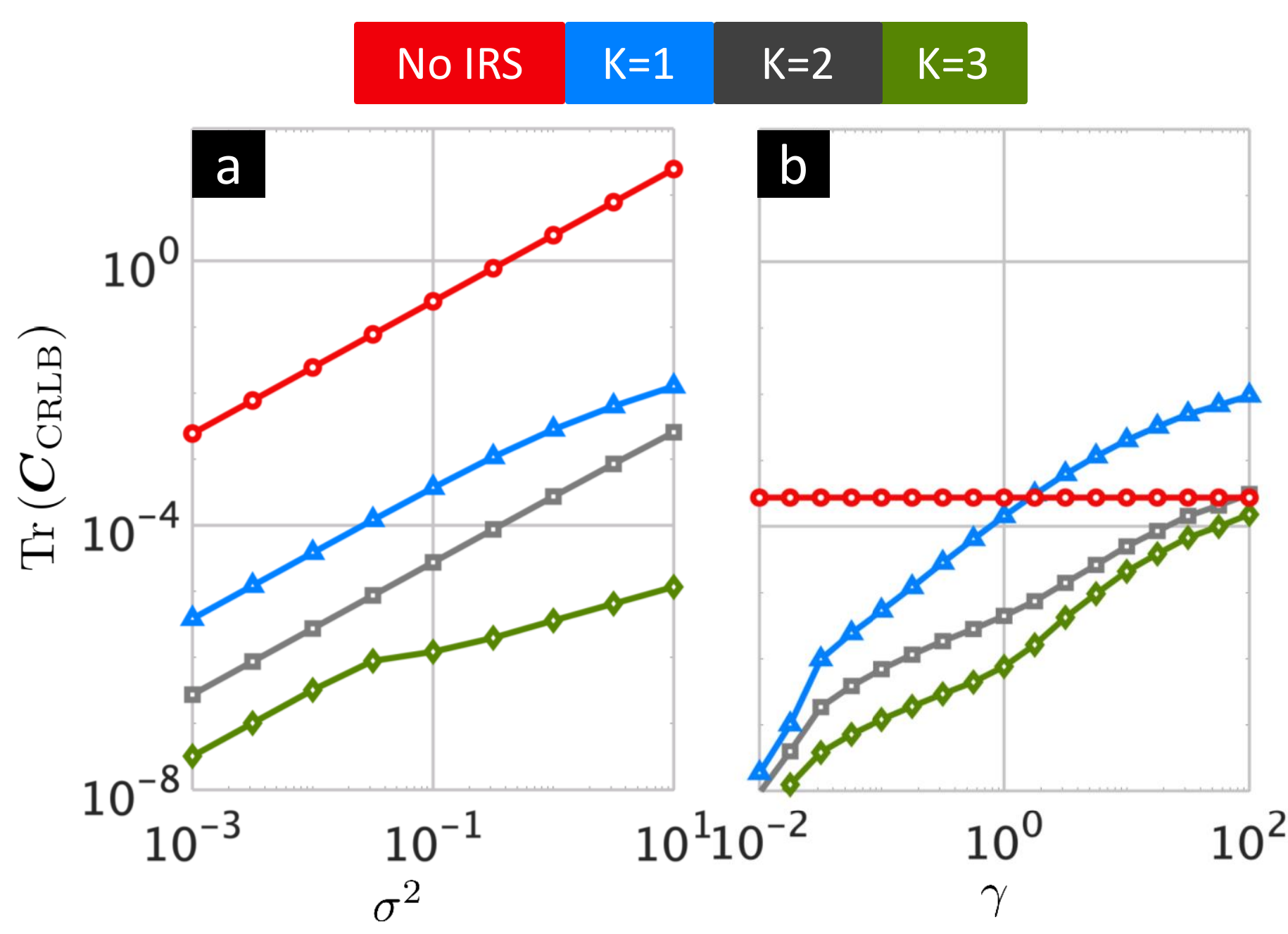}
	\caption{The CRLB of the target back-scattering coefficients and Doppler shifts  for different values of  (a) $\sigma^2\in [10^{-3},10^{1}]$ and  $\gamma=0.1$, and (b)  $\gamma\in [10^{-2},10^{2}]$ and \ze{$\sigma^2=0.1$}.}
	\label{fig::result}
\end{figure}
%---------------------------------
Throughout all numerical experiments, the radar and target were positioned at, respectively, [$0$ m, $0$ m] and [$0$ m, $5000$ m] in the 2-D Cartesian plane. The target's normalized Doppler shifts were  drawn from the uniform distribution in the interval [0.1,0.3) with mean 0.2.
%\ze{The target's normalized Doppler shifts were  generated such that $\bnu \sim \mathcal{N}\left(\bmu_{\bnu},\sigma^2_{\bnu}\bI_{_{K+1}}\right)$}.
We considered the  distance dependent LoS  path loss $h_{_{\LoS}}=L_0(\frac{d}{d_0})^{-\beta_0}$, where $L_0=-30$ dB is the path loss at the reference distance $d_0=1$ m and the path-loss exponent is set as $\beta_0=2.5$. The  NLoS path loss is set proportionally according to $\emph{LoS-to-NLoS LSR}$ $\gamma$. % in~\eqref{eq::gamma}.
%In all our experiments we used  constant Lagrangian multipliers  %$\eta_k=0.1$ for $k \in \{1,\ldots,K\}$. 
We positioned a single IRS platform in  [$2500$ m, $2500$ m]. Next, we positioned the second and third IRS platforms in  [-$2500$ m, $2500$ m] and [$0$ m, $2500$ m], respectively. For a point like target, given the  radar, target and IRS  positions,  the corresponding  radar--IRS$_{k}$ and  target--IRS$_k$ angles $\theta_{ir,k}$ and $\theta_{ti,k}$, for $k \in \{1,\ldots,K\}$ are calculated geometrically in the 2-D plane. In all our experiments  in order to control the $\emph{LoS-to-NLoS LSR}$ $\gamma$, the complex  target reflectivity  coefficients in $\balpha$ were  sampled from 
%\ze{the complex  target reflectivity  coefficients in $\balpha$ were  sampled from $\balpha \sim \mathcal{CN}\left(\bmu_{\balpha},\sigma^2_{\balpha}\bI_{_{K+1}}\right)$}
independent circularly symmetric complex Gaussian distribution with zero mean and variance of unity  and  scaled such that we have $|\alpha_{_0} h_{_{\LoS}}|^2=\gamma$ and $\sum_{k=1}^{K}|\alpha_k h_{_{\NLoS,k}}|^2=1$. 

For fixed $\gamma=0.1$, Fig.~\ref{fig::result}a shows  the  CRLB  versus the noise  variance,  achieved  using the Doppler-aware phase shifts for IRS  designed in this paper. Moreover,  the effectiveness of  deploying  multiple IRS platforms  versus the no-IRS and single-IRS  deployment is  illustrated. Fig.~\ref{fig::result}b compares the performance for different  %$\emph{LoS-to-NLoS SNR}$ of 
$\gamma\in [10^{-2},10^{2}]$.  We  observe that IRS deployment is effective  in  improving estimation accuracy when $\gamma$  decreases to values  below $1$, whereas deploying  more than  one IRS  can improve the estimation accuracy for all $\emph{LoS-to-NLoS LSR}$ values. Fig.~\ref{fig::result}a-b indicate that deploying multiple IRS platforms  in a radar system with the optimized  Doppler-aware phase shifts is beneficial in  target estimation. 
\section{Conclusion}
\label{sec:conclusion}
Drawing on the recent advances in IRS-aided communications and radar, we investigated a pulse-Doppler multi-IRS-aided radar system with the goal of improving the estimation accuracy of a moving target. We carried out the theoretical estimation accuracy based on Fisher information analysis, aimed at designing the Doppler-aware  phase shifts.  %Our  proposed  radar system  equipped with multiple IRSs to impose the Doppler-resilient phase-shifts on the impinging signal, 
Our approach improves the estimation accuracy of the Doppler shifts and target  complex reflectivity %. We  illustrated that deploying multiple  IRS platforms  versus a single IRS can  further improve the estimation accuracy. 
over a single IRS-aided radar. %This performance analysis is a precursor to developing methods for optimal IRS placement.
% \textcolor{red}{rewrite this section. you are just repeating the Intro and abstract here. Discuss some related works instead and connect them with our work. For example, see how I write the summary in my paper: \url{https://ieeexplore.ieee.org/abstract/document/9858030} } 
% In this work, the backscattered signal from a point-like target  with known location  is used to carry out a  aimed at designing the  phase-shifts and  investigating  the theoretical  accuracy  achievable in the estimation of  target parameters in IRS-aided radar. 
%Our approach used the   CRLB as a lower bound for any unbiased estimator as a benchmark for  target estimation.
%\textcolor{blue}{Furthermore, the joint target parameter estimation problem from the lens of the Cram\'er Rao lower bound (CRLB) is investigated. }
%\appendices
%\textcolor{red}{No need for an appendix in a letter, unless a reviewer suggests it. Appendices are appropriate for long Transactions papers}
%\section{Detailed Derivations for FIM}\label{sec:fisher}
%\label{sec_appA}
%\clearpage
%\balance
\bibliographystyle{IEEEtran}
\bibliography{refs}

\end{document}